# Coherent Contributions to Population Dynamics in a Semiconductor Microcavity

J. Paul,[1,2] H. Rose,[3] E. Swagel,[2,4] T. Meier,[3] J. K. Wahlstrand,[5] and A. D. Bristow,[1,2,*]

[1]*Department of Physics and Astronomy, West Virginia University, Morgantown, WV 26506-6315, USA*

[2]*Associate, Nanoscale Device Characterization Division, National Institute of Standards and Technology, Gaithersburg, MD 20899, USA*

[3]*Paderborn University, Department of Physics, Warburger Strasse 100, D-33098 Paderborn, Germany*

[4]*Department of Physics, Brown University, RI 02912, USA*

[5]*Nanoscale Device Characterization Division, National Institute of Standards and Technology, Gaithersburg, MD 20899, USA*

**E-mail: alan.bristow@mail.wvu.edu*

**Abstract:** Multidimensional coherent spectroscopy (MDCS) is used to separate coherent and incoherent nonlinear contributions to the population-time dynamics in a GaAs-based semiconductor microcavity encapsulating a single InGaAs quantum well. In a three-pulse four-wave-mixing scheme, the second delay time is the population time that in MDCS probes excited-state coherences and population dynamics. Nonlinear optical interactions can mix these contributions, which are isolated here for the lower- and upper-exciton-polariton through the self- and mutual-interaction features. Results show fast decays and oscillations arising from the coherent response, including a broad stripe along the absorption energy axis, and longer time mutual-interaction features that do not obey a simple population decay model. These results are qualitatively replicated by Bloch equation simulations for the 1s exciton strongly coupled to the intracavity field. The simulations allow for separation of coherent and incoherent Pauli-blocking and Coulomb interaction terms within the $\chi^{(3)}$-limit, and direct comparison of each feature in one-quantum rephasing and zero-quantum spectra.

## I. INTRODUCTION

Semiconductor microcavities have been an ideal system for exploring coherent light-matter interaction phenomena, such as enhanced nanolasing [1,2], strong coupling between photonic and electronic modes, and even Bose-Einstein condensation [3,4]. The microcavity increases the optical density of states at specific photon energies [5]. For planar semiconductor microcavity systems based on III-V technology, the cavity mirrors are typically distributed Bragg reflectors (DBR) encapsulating a discrete absorbing state such as a quantum well (QW) at the antinode of the cavity [6–8]. If the cavity mode is tuned close to the excitonic mode of the QW, the eigenstates mix and normal-mode splitting produces lower- (LP) and upper-exciton-polariton (UP) branches that exhibit avoided-crossing behavior in the wavevector or cavity detuning dependence and a trap-like LP dispersion as a function of incident angle [9,10]. The strength of the coupling depends on the detuning of the cavity with respect to the exciton and the vacuum Rabi splitting, which encompasses cavity loss and impedance matching of the photonic and electronic modes [11]. Planar microcavities exhibit strong incident angle dependence and parametric scattering mechanisms [12,13] can result in a dense polariton population at low wavevectors [14–18]. These phenomena are a few examples of the rich nonlinear optical processes made possible by strong coupling of the photonic and electronic modes [19–27]. As such, nonlinear optics of semiconductor microcavities remain fertile ground for exploring novel coherent phenomena [28–32] and applications in quantum information processing [33–38].

Multidimensional coherent spectroscopy (MDCS), based on four-wave mixing, unfolds complex spectra to isolate various contributions to the nonlinear optical response [39,40]. MDCS has been applied to GaAs-based semiconductor microcavities in recent years [41-45], where it has tracked the relative strengths and linewidths of the self- and mutual-interaction spectral features as a function of detuning through the anti-crossing [41], showing that near zero detuning the LP and UP modes experience similar spectral properties due to strong hybridization. MDCS spectra have been interpreted through double-sided Feynman diagrams linking specific features to Liouville-space quantum pathways and formalizing them within the Gross-Pitaevskii equation [42]. It has been shown that coherent contributions to the nonlinear response are consistent with semiconductor many-body interactions, which MDCS isolated through selection rules and higher-order excitation pulse sequences [43]. Furthermore, higher-order MDCS spectra have been shown to correspond to the Tavis-Cummings ladder of excited quantum states and those states follow the polariton dispersion [44]. However, the third-order response exhibited in MDCS arises from excited-state population and coherent contributions that have not been isolated in microcavity samples. Taking full advantage of the information contained in MDCS spectra often requires comparison to theoretical calculations that include many-body effects [45].

Excited-state population dynamics and coherences can be followed using three-pulse MDCS, varying the population time between the second and third pulses [46–49]. In this paper, we report measurements of MDCS spectra of a semiconductor microcavity as a function of population time [41]. Analysis of the exciton-polaritons from the one-quantum rephasing spectra reveals two-exponential decays for the diagonal (self-interaction) and off-diagonal (mutual-interaction) features, and oscillations in the mutual-interaction features. Zero-quantum spectra confirm non-radiative coherences between excited lower- and upper-polariton branches, consistent with the observed oscillations. Results are compared to numerical simulations of Bloch equations extended to include microcavity effects [28] and optical nonlinearities [50]. Both Pauli blocking and Coulomb interactions contribute to the coherent and incoherent population dynamics, agreeing well with the experiments despite neglecting inhomogeneity and the dispersion of polariton modes.

## II. METHODS

### *II.A Experiment*

A monolithic, planar semiconductor microcavity was grown by molecular beam epitaxy on a GaAs substrate. As illustrated in Fig. 1(a), its structure consists of two GaAs/AlAs (14.5 and 12 bilayer) distributed Bragg reflectors to produce a wedged, λ cavity of GaAs with a single 8-nm, $In_{0.04}Ga_{0.96}As$ QW positioned at the cavity's antinode [9]. The structure gives rise to an exciton eigenmode, $E_X = 1490.1$ meV (832 nm) at temperatures near 6 K, and a tunable cavity mode, $E_\gamma(\Delta)$, where the cavity detuning is $\Delta = E_\gamma - E_X$. Near $\Delta = 0$ meV, LP ($E_-$) and UP ($E_+$) branches appear in the linear absorbance spectra and imitate a three-level system, as shown in Fig. 1(b). The $\Delta$ dependence of the polariton branches have energies $E_\pm(\Delta) = \frac{1}{2}\left[2E_X+\Delta\pm\left(\Delta^2+E_{\mathrm{VRS}}^2\right)^{1/2}\right]$, from which a vacuum Rabi splitting $E_{\mathrm{VRS}} \approx 3.1$ meV is extracted; see Fig. 1(c). The cavity mode is incident-angle ($\theta$) dependent, giving $E_\pm(\theta) = \frac{1}{2}\Delta(\theta) \pm \sqrt{4E_{\mathrm{VRS}}^2 + \Delta(\theta)^2}$, which is modelled in Fig. 1(d) for $\Delta = -2$ meV.

Measurements are performed with a multidimensional optical nonlinear spectrometer (MONSTR) [51] with 120-fs pulses from a single 76-MHz mode-locked Ti:sapphire laser oscillator. Pulses have a spectral bandwidth of 15 meV, tuned between the LP and UP features; see the inset of Fig. 1(b). Three excitation pulses are incident on the microcavity sample in a non-collinear box geometry, as shown in Fig. 1(a), with the laser tuned to excite LP and UP equally. Using a single 25-cm focusing lens, the beams impinge the sample with an incident angle of $\theta_i \approx 5^{\text{o}}$ or in-plane wavevectors of $k_{||,i} = (E_\gamma/\hbar c)\sin\theta \approx 0.68$ μm$^{-1}$, as illustrated in Fig. 1(d). Incident pulses A, B and C each have a fluence of 52 nJ/cm$^2$ with a $1/e^2$-diameter of 15 μm, equivalent to an excitation density of $\approx 10^{11}$ cm$^{-2}$.

Along with a local oscillator pulse derived from the laser, four-wave mixing (FWM) emission is collected in the $-k_{\text{A}} + k_{\text{B}} + k_{\text{C}}$ direction and detected in an imaging spectrometer with a charged-coupled device. Spectral interferograms $S_{\text{I}}(\omega_t)$ capture the emission time axis $(t)$ defined by pulse C, for fixed evolution time $(\tau)$ between pulses A and B, and fixed population time ($T$) between pulses B and C. Spectral interferograms are recorded for a range of $\tau$ to produce rephasing two-dimensional spectra $S_{\text{I}}(-\omega_\tau, T, \omega_t)$ by a numerical transform with respect to $\tau$.

Various polarization configurations were measured using automated variable retarders [52], but only co-circular $(\sigma^+\sigma^+\sigma^+\sigma^+)$ spectra are presented here. This polarization configuration forbids excitation of biexciton-like states through the excitonic selection rules, maintaining the simple level scheme in Fig. 1(b) [53]. Zero-quantum spectra, $S_{\text{I}}(\tau, \omega_T, \omega_t)$, are measured by scanning the population time ($T$) [46] and computed using a linear prediction singular value decomposition routine [54]. Spectral quality is improved by using variable retarders to perform phase cycling at each step of the scans [55].

*II.B Theory*

To numerically simulate the optical response of the coupled QW-microcavity system, we consider a set of Bloch equations that describe the polarization dynamics of the excitonic transitions $p$ in the presence of an intracavity field $\alpha$. These equations contain the Coulomb-interaction on a Hartree-Fock level and the incoherent excitonic occupations $\bar{N}$, in the form:

$$\partial_t \alpha = (-\gamma_\alpha - i\omega_\alpha)\alpha + igp + \tilde{g}E(t), \tag{1a}$$

$$\partial_t p = \left(-\gamma_p - i\omega_p\right)p + ig\alpha - ib\alpha(|p|^2 + \bar{N}) - iVp(|p|^2 + \bar{N}), \tag{1b}$$

$$\partial_t \bar{N} = -\gamma_N \bar{N} - \left(\gamma_N - 2\gamma_p\right)pp^*, \tag{1c}$$

where $g$ denotes the light-matter coupling strength and $\tilde{g}$ the coupling from the external field $E(t)$ into the cavity. The exciton resonance frequency is given by $\omega_p$, $\omega_\alpha$ is the eigenfrequency of the energetically lowest cavity mode, $\gamma_p$ and $\gamma_\alpha$ are the inverse of the exciton dephasing time and cavity photon lifetime respectively, and $\gamma_N$ is the inverse of the incoherent exciton population lifetime. The parameters $b$ and $V$ denote the optical nonlinearities that give rise to the FWM signal, where $b$ denotes Pauli blocking and $V$ the Coulomb interaction between excitons of the same spin.

Equations (1a)-(1c) can be obtained by considering a Hamiltonian describing a two-band model, including band energies, Coulomb interaction, and the light-matter interaction. The time dynamics of the operator $\hat{A}$ is computed with the Heisenberg equation of motion [56],

$$\frac{\partial}{\partial t}\hat{A} = \frac{1}{i\hbar}\left[\hat{A},\hat{H}\right]_-, \tag{2}$$

where the brackets denote the commutator $\left[\hat{A},\hat{H}\right]_- = \hat{A}\hat{H} - \hat{H}\hat{A}$.

The Coulomb interaction gives rise to a hierarchy problem that is truncated with the dynamics-controlled truncation (DCT) scheme, which assumes that excitation contributions can be attributed to a power of the electric field [50,57], with exciton occupation and coherent terms [55–57]. Hence, the excitonic occupation $N$ is decomposed into a coherent ($pp^*$) and an incoherent ($\bar{N}$) part via [58–60]

$$N = \bar{N} + pp^*, \tag{3}$$

which is then used in Eqs. (1a)-(1c). The resulting quantities are projected onto the energetically lowest single-exciton and two-exciton states, which models the excitation of the spectrally isolated 1s-exciton [61].

The external electric field that drives the QW-microcavity system is modelled as a sum of Gaussians

$$E(t) = \sum_{i=1}^{3} E_{0,i} e^{-2\ln 2[(t-t_i)/t_w]^2} \mathrm{e}^{-i\omega_L t} + c.c. \tag{4}$$

where the amplitude of pulse $i$ is $E_{0,i}$, the arrival time of pulse $i$ is $t_i$, the full-width at half-maximum is $t_w$, and the central frequency is $\omega_L$. Optical selection rules match experiments, namely only a single spin state of the 1s-exciton is accessed with $\sigma^+$-polarized light, suppressing bound biexciton contributions.

These equations are solved perturbatively within the rotating-wave approximation [62] up to third-order in the external field, taking into account the phase-matching condition. The FWM signal is obtained by projecting the third-order intracavity field $\boldsymbol{\alpha}^{(3)}(t,T,\tau)$, which is Fourier transformed into different spectral spaces. Rephasing one-quantum spectra are given by [63]

$$S_{\mathrm{I}}(-\omega_\tau,T,\omega_t)=\int_{-\infty}^{\infty}\int_{-\infty}^{\infty}\boldsymbol{\alpha}^{(3)}(t,T,\tau)\,\boldsymbol{\epsilon}e^{i\omega_t(t)+i(\omega_T)T+i(-\omega_\tau)\tau}dtd\tau \tag{5}$$

and zero-quantum spectra by

$$S_{\mathrm{I}}(\tau,\omega_T,\omega_t)=\int_{-\infty}^{\infty}\int_{-\infty}^{\infty}\boldsymbol{\alpha}^{(3)}(t,T,\tau)\,\boldsymbol{\epsilon}e^{i\omega_t(t)+i(\omega_T)T+i(-\omega_\tau)\tau}dtdT \tag{6}$$

where $\boldsymbol{\epsilon}$ is the $\sigma^+$-polarized detection. The optical nonlinearities are adjusted to give good agreement with the experiment, which led to the choice of $\hbar b=2$ meV and $\hbar V=1$ meV for all simulations presented in this paper. Note that since the magnitude of the experimental spectra are not measured in absolute units, it is the ratio of these values that is important, rather than the absolute magnitudes of each.

## III. EXPERIMENTAL RESULTS

Figure 2 shows the amplitude [(a)-(f)] and real-part [(g)-(l)] of the experimental one-quantum rephasing spectra $S_{\mathrm{I}}(-\omega_\tau,T,\omega_t)$ for the semiconductor microcavity at $\Delta=0$ meV excited with co-circular polarization and for a selection of population times. For all spectra, self-interaction features on the diagonal (LP and UP) and mutual-interaction features off of the diagonal (UP-LP and LP-UP) dominate [41,52,64,65]. The features exhibit mild inhomogeneity [66], most readily seen as a slight elongation of the LP mode along the diagonal direction of the spectrum (top-left to bottom-right). At $\Delta=0$ meV, the degree of inhomogeneous broadening can be characterized as $(\sigma/\gamma)_{LP}=1.12\pm0.08$ and $(\sigma/\gamma)_{UP}=0.81\pm0.08$, where $\gamma$ and $\sigma$ correspond to the homogeneous and inhomogeneous broadening linewidths respectively [67] and within error match those previously reported for this sample [41]. While these values are not zero, they are not as large as $(\sigma/\gamma)_{QW}>3$ [66] for bare quantum wells or even the degree of inhomogeneous broadening reported for this sample at detuning values further away from $\Delta=0$ meV [41]. Values of $(\sigma/\gamma)\approx1$ for all $\Delta$ values used in experiments reported here.

The real-part of the spectra show dispersive lineshapes. The cross-diagonal spectral phase of the LP and UP features is opposite, and the off-diagonal phases are even more complicated, but all exhibit consistent spectral phase stability over the range of population times. These spectral phase properties are different from purely excitonic systems, where heavy- and light-hole excitons typically have similar spectral phase [53,68] with a $T$ dependence [48].

Dashed construction lines in Fig. 2(a-c) indicate cross-diagonal line slices through the four features, which are shown in Fig. 2(m-o) as a function of increasing $T$. The diagonal line slices exhibit Voigt profiles. At small $T$, the strongest feature is the LP mode, while at larger $T$ it is the LP-UP feature. This behavior is typically ascribed to incoherent downhill population transfer from the upper to lower excited states in multi-level systems, namely absorption into UP and relaxation to LP, then emission. The signal at short times is also enhanced, which in pump-probe measurements is referred to as the coherent artifact/transient and can be associated with many-body excitonic interactions [69]. A reduction in signal strength is seen for all four peaks with increasing $T$, although the decay rates are not identical. The slightly slower decay of LP-UP compared to the diagonal features is consistent with downhill population transfer. However, the relatively strong UP-LP feature at large $T$ is inconsistent with this reasoning and requires a longer-lived contribution to mix the exciton-polariton populations.

In addition to the dominant features of the rephasing spectra, there are weaker vertical stripes along the absorption axis ($-\hbar\omega_\tau$). Experimentally these stripes are caused by increased FWM emission at $\tau = 0$, when pulses A and B overlap in time, which results in a broad pedestal in the absorption frequency dimension. In Fig. 2, the stripes are particularly noticeable at the LP emission energy, possibly due to an excitation-induced dephasing process that increases spectral weight at the lowest energy emission state [70]. The vertical stripes are not seen for bare quantum wells [45] nor do they correspond to wings in the linear absorption spectrum of the microcavity [see inset of Figure 1(b)], and therefore seem to be a consequence of nonlinear interaction in the cavity.

Figure 3 (a) – (d) show the extracted peak amplitudes of the dominant features as a function of $T$ for three different detuning values. The LP and UP amplitude time dependence is empirically fit by a double exponential decay. Fast ($\approx$1.1 ps) and slow ($\approx$50 ps or longer for some features) components are consistent with a coherent transient and incoherent population, respectively. The off-diagonal (LP-UP, UP-LP) features decay on similar time scales but also exhibit oscillations indicative of excited-state coherences; see Fig. 3(c)-(d) and insets for oscillations at detuning values $\Delta = -2.2$, 0, 2.2 meV. The oscillation frequencies match the splitting of the two polariton branches for each $\Delta$. Also, the oscillation damping time is approximately double that of the fast decay of the diagonal features.

Figure 3 (e) and (f) show zero-quantum spectra, $|S_{\mathrm{I}}(\tau, \omega_T, \omega_t)|$, for $\Delta = 2.2$ meV and $-2.2$ meV. For these spectra, $\tau$ is fixed at 1.5 ps to reduce overlap of the cavity field arising purely from

the first excitation pulse with subsequent pulses. The spectra isolate the non-radiative, excited-state coherence as features labelled LP-Δ and Δ+UP, which are associated with the oscillations of the rephasing LP-UP and UP-LP features respectively. Spectra at these two Δ values are sufficient to confirm that the oscillations match the splitting of the LP and UP branches. For positive detuning the coherent coupling between the excited LP and UP states is strong, resulting in off-zero features of near equal strength. For negative detuning, the strength of the off-zero features is asymmetric, favoring emission from the (more cavity-like) LP mode and which is equivalent to the Stokes line if this process were Raman scattering.

The degree of correlation of fluctuations of the LP and UP transitions is given by $R_{LP-UP(UP-LP)} = \left[\gamma_{LP} + \gamma_{UP} - \gamma_{LP-UP(UP-LP)}\right]/2(\gamma_{LP}\gamma_{UP})^{1/2}$ [71,72], where $\gamma_{LP(UP)}$ denotes the dephasing rates of the LP(UP) feature determined from the cross-diagonal linewidths of the self-interaction features in the rephasing spectra and $\gamma_{LP-UP(UP-LP)}$ denotes the dephasing rates (or width along $\hbar\omega_T$) of the off-zero features in the zero-quantum spectra [73]. We estimate a 20% uncertainty in finding the dephasing rates from fits of our spectra. Calculated for $\Delta = 0$ meV, the degree of correlations are $R_{\mathrm{LP-UP}} = -0.09 \pm 0.26$ and $R_{\mathrm{UP-LP}} = 0.29 \pm 0.24$. Values calculated from the spectra at the other detuning positions measured show similar results within uncertainties. These values are effectively equal within error and indicate that fluctuations in the LP and UP states are mostly uncorrelated during the population time, since $R = 1, 0$ and $-1$ correspond to perfectly correlated, uncorrelated and anticorrelated states respectively. These results differ from the anticorrelated fluctuations observed between heavy- and light-hole excitons [72] or the partly correlated fluctuations observed in potassium vapor [73], but better match the low-temperature uncorrelated fluctuations observed of an exciton-biexciton ladder at low-temperature [71]. For the latter system, excitons and biexcitons have a large energy difference and scattering occurs by phonons with different dispersions. In contrast, the LP and UP modes of the microcavity are closer in energy but their dispersions are significantly different.

## IV. SIMULATION RESULTS

Simulations of modified Bloch equations were performed as described in section IIB, using the rotating wave approximation with only a difference between the exciton and cavity modes. Spectra were simulated using experimental parameters, using $E_X = 1491$ meV, $E_\gamma = E_X + \Delta$, $\hbar\omega_L = E_X + \Delta/2$ and the amplitude is derived from using 100-fs pulses. Additional fixed parameters include $g = 1.65$ meV and $\tilde{g} = 1$. Spectra were simulated using artificially shortened lifetimes to compensate for the lack of inhomogeneity and dispersion in the model [$1/\gamma_N = 46.4$ ps (population), $1/\gamma_p = 1.67$ ps (polarization) and $1/\gamma_\alpha = 1.38$ ps (cavity)]. Nonlinear terms $b$ and $V$ were the only free parameters in the calculations.

Figure 4 shows the full calculation (e,j) of the co-circularly polarized rephasing spectrum for $T = 0.5$ ps and $\Delta = 0$ meV. Separate coherent and incoherent contributions of the (a,b,f,g) Pauli-blocking and (c,d,h,i) Coulomb nonlinear source terms combine linearly to form the full calculation. All spectra are normalized to their strongest feature. The full calculation spectrum shows good qualitative agreement with experiment: There are four dominant features, namely two diagonal (off-diagonal) self-interaction (mutual-interaction) terms. The relative amplitude matches the experiment and the real parts of the full calculation show that the LP and UP have opposite spectral phase. Additionally, there are vertical ($\hbar\omega_\tau$-axis) stripes located at the two emission energies, with the strongest stripe at the LP emission energy.

The global phase of the full simulation is matched to the experiment, but the relative spectral phase of each nonlinear contribution is unique. Consequently, they combine with different relative strengths to create the full calculation. The Pauli and incoherent Coulomb contributions show stronger diagonal features, whereas the coherent Coulomb interactions show slightly stronger off-diagonal features. Even at this short population time, a transfer of spectral weight from the UP to LP states is mediated by Coulomb interactions, akin to processes seen for continuum states in excited bulk semiconductors [70] which is empirically attributed to excitation-induced dephasing [70].

Figure 5 (a)-(f) shows calculated spectra for a range of population times matching those shown in Fig. 2. As with experiment, the UP-LP feature does not vanish at large $T$, confirming strong mixing between the lower and upper polariton branches throughout the time range presented and that this system deviates from a purely excitonic system [28]. Coherent and incoherent Pauli and Coulomb contributions to each spectral feature are extracted from a more complete series of rephasing simulations and plotted versus $T$ in Fig. 6 (a)-(d) for $\Delta = 0$ meV. Each contribution produces a peak with complex amplitude, so we plot both real and imaginary parts. To match the observed decays in Fig. 3 (a-d) rather than the spectra, slightly longer decay times were used (namely, $1/\gamma_N = 92.2$ ps, $1/\gamma_p = 3.32$ ps, and $1/\gamma_\alpha = 2.76$ ps). As with the experiment, fast and slow decays are exhibited by the self-interaction features, with additional oscillations in the mutual-interaction features. The fast response is strong and arises from a combination of coherent Pauli and Coulomb contributions and is proportional to $\exp[-(\gamma_p + \gamma_\alpha)T]$, whereas the slower response is a result of the incoherent contributions and proportional to $\exp(-\gamma_N T)$. Overall, the polariton decay rate is $^1/_2\,(\gamma_p + \gamma_\alpha)$, where $\gamma_\alpha \approx \gamma_p$, so the fast decay is similar to $\sim 2\gamma_p$, or the coherent limit in a purely electronic two-level system without an optical cavity. However, the behavior of this system is different because the long-lived intracavity field has a direct impact on both coherent occupation $pp^*$ and incoherent occupation $\bar{N}$. Fig. 6 (a)-(d) demonstrate that $\bar{N}$ shows coherent-like effects at short times, because it has a component with a similar decay time to the coherent contributions. Hence, this analysis assists in separating coherent and incoherent contributions.

For direct comparison to the experimental transients in Fig. 3, the *T*-dependent amplitude of each spectral feature extracted from the series of full calculations is plotted in Fig. 6 (e)-(h). Overall, the simulated $T$ dependence agrees with the experimental results, showing several prominent things: (i) The incoherent Pauli-blocking contributions give rise to the strong anisotropy of the off-diagonal features, consistent with downhill population transfer from the upper to lower excited states of a three-level system. (ii) The incoherent Coulomb interaction produces long-lived off-diagonal features that are nearly identical in magnitude and are therefore responsible for the contributions to the long-lived LP-UP feature of the rephasing spectra. (iii) Pauli and Coulomb contributions exhibit oscillations during the first few picoseconds of the population time.

Simulations of zero-quantum spectra confirm the existence of off-zero features arising from nonlinear contributions; see Fig. 7 for detuning values that match experiment. Interestingly, the oscillations in the population time are not limited to the coherent contributions, suggesting that energy and/or population *slosh* between the two excited states during these short times. However, the difference in the relative strengths of the off-zero features in experiment and theory are significant. The experiment shows much larger anisotropy in the amplitude of the off-zero features than the theory at –2.2 meV. This is most likely a result of strongly dispersive polariton states in the microcavity and more complex nonlinear terms. Accurate replication of the zero-quantum features begs inclusion of a full semiconductor Bloch treatment beyond the Hartree-Fock level for the exciton states and the angle dependence of the photonic coupling.

## V. CONCLUSION

In summary, multidimensional coherent spectra of a semiconductor microcavity for a range of population times provide insight into the population dynamics and comparison to simulations allows for separation of coherent and incoherent contributions to the nonlinearity.

Good qualitative agreement was found between the theory and experiment using only two parameters ($\hbar b = 2$ meV for Pauli blocking and $\hbar V = 1$ meV for Coulomb interaction between excitons of the same spin). Both measured and calculated spectra showed fast and slow decays of all peaks, oscillations in the off-diagonal peaks, and vertical stripes. Coherent contributions dominate the short times and incoherent contributions dominate the longer times, however, the separation is not completely clear, since the incoherent response also exhibits a fast decay component at early times, which is most likely due to the cavity field duration. Interestingly, lower- and upper-polariton peaks exhibit opposite spectral phase, quite unlike purely excitonic systems. Moreover, the system exhibits excited-state oscillations and (partial) energy-transfer processes that are both replicated by the theory.

This study focuses on the relative importance of the nonlinear interactions in an monolithic semiconductor microcavity based on III-V technology, however this combined simulation and coherent spectroscopy approach would be a powerful tool for analyzing coherent and incoherent population dynamics in photonic or polaritonic systems with different sources of nonlinearity [64,74–76].

**FUNDING**

This work was supported by NIST Award # 70NANB18H238, Deutsche Forschungsgemeinschaft (DFG) Collaborative Research Center TRR 142 (project 231447078, project A02) and a grant for computing time from PC2 (Paderborn Center for Parallel Computing).

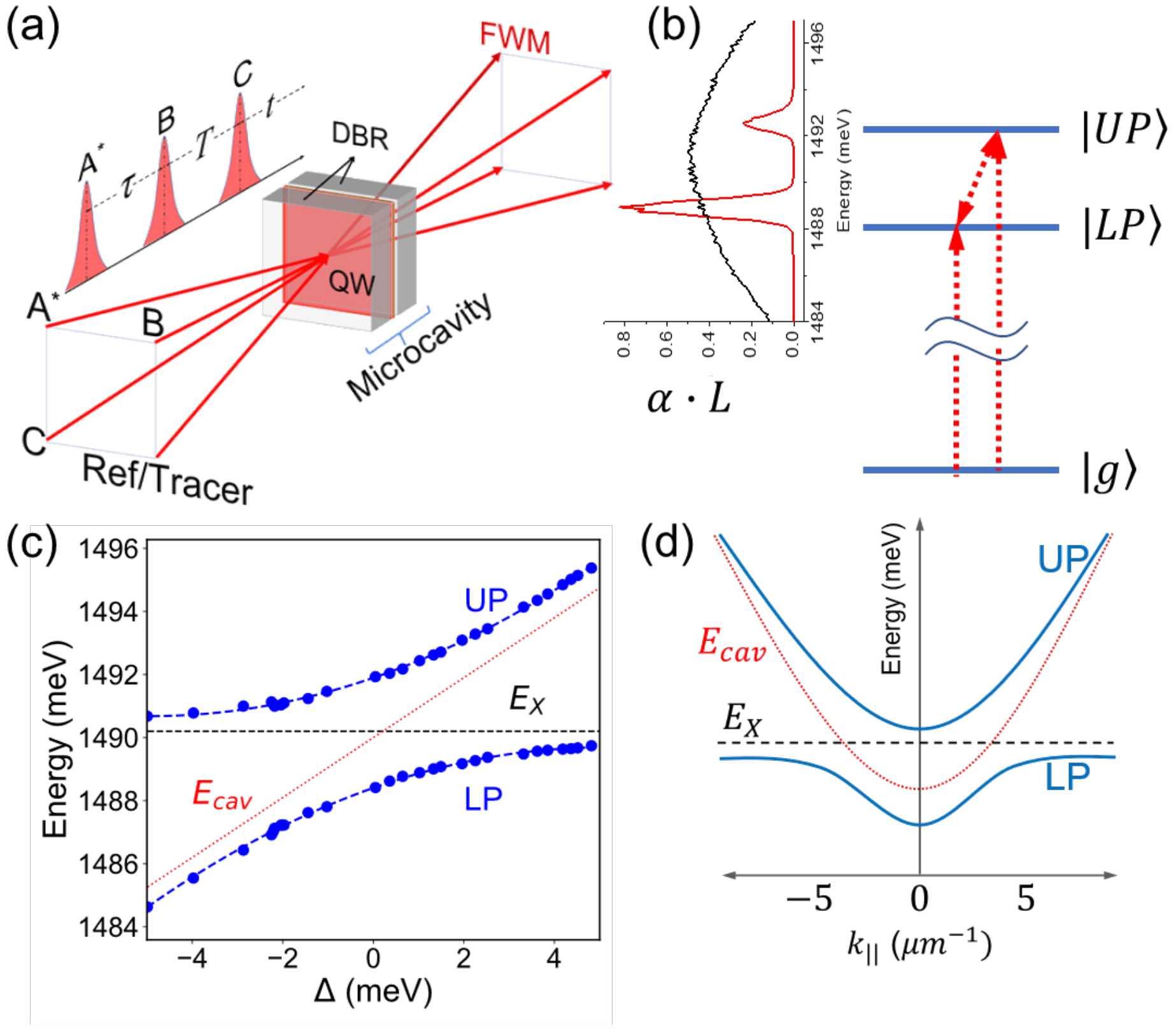

**Fig. 1**. (a) Experimental setup for three-pulse coherent spectroscopy of a monolithic GaAs-based semiconductor microcavity. Complex four-wave mixing emission spectra are measured as a function of evolution time $\tau$ and population time $T$. (b) Three-level energy scheme for the exciton-polaritons. The inset shows the microcavity absorbance ($\alpha{\cdot}L$) with the laser spectrum overlaid in arbitrary units. (c) Detuning ($\Delta$) dependence of the exciton-polariton branches showing the extracted cavity and exciton positions at ≈6 K. (d) Calculated in-plane wavevector ($k_{\|}$) dependence of the exciton-polaritons based on the extracted energy positions for $\Delta = -2$ meV.

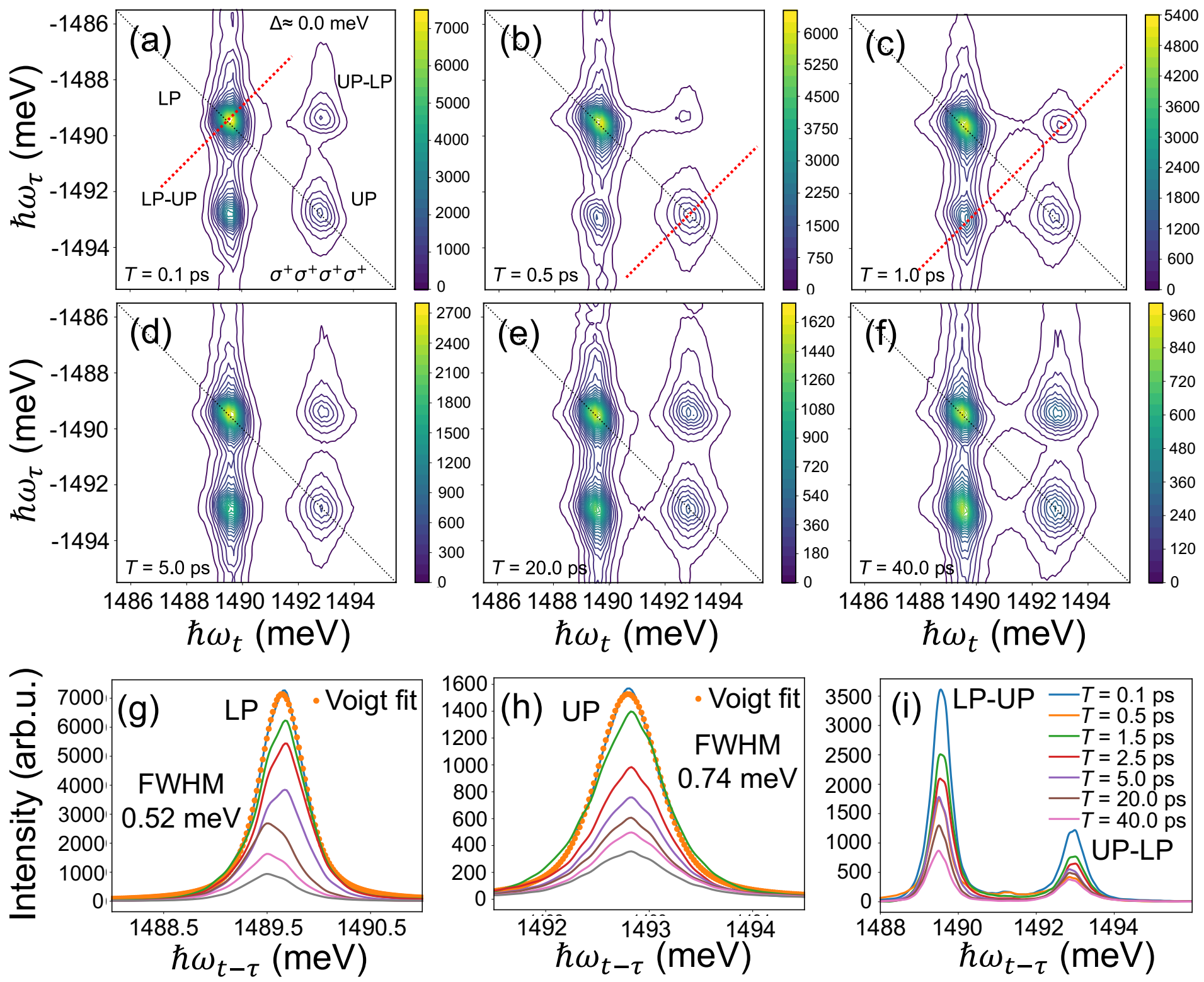

**Fig. 2**. (a) – (f) Two-dimensional rephasing spectra, $|S_{\mathrm{I}}(-\omega_\tau, T, \omega_t)|$, of the semiconductor microcavity for zero detuning ($\Delta$ = 0 meV), co-circular polarization ($\sigma^+\sigma^+\sigma^+\sigma^+$) and a range of population times $0.1 \leq T \leq 40$ ps. (g-l) Real part spectra $\mathrm{Re}[S_{\mathrm{I}}(-\omega_\tau, T, \omega_t)]$ corresponding to (a)-(f). One-dimensional cross-diagonal line slices through the rephasing spectra for the diagonal (m) lower-polariton (n) upper-polariton and (o) off-diagonal features.

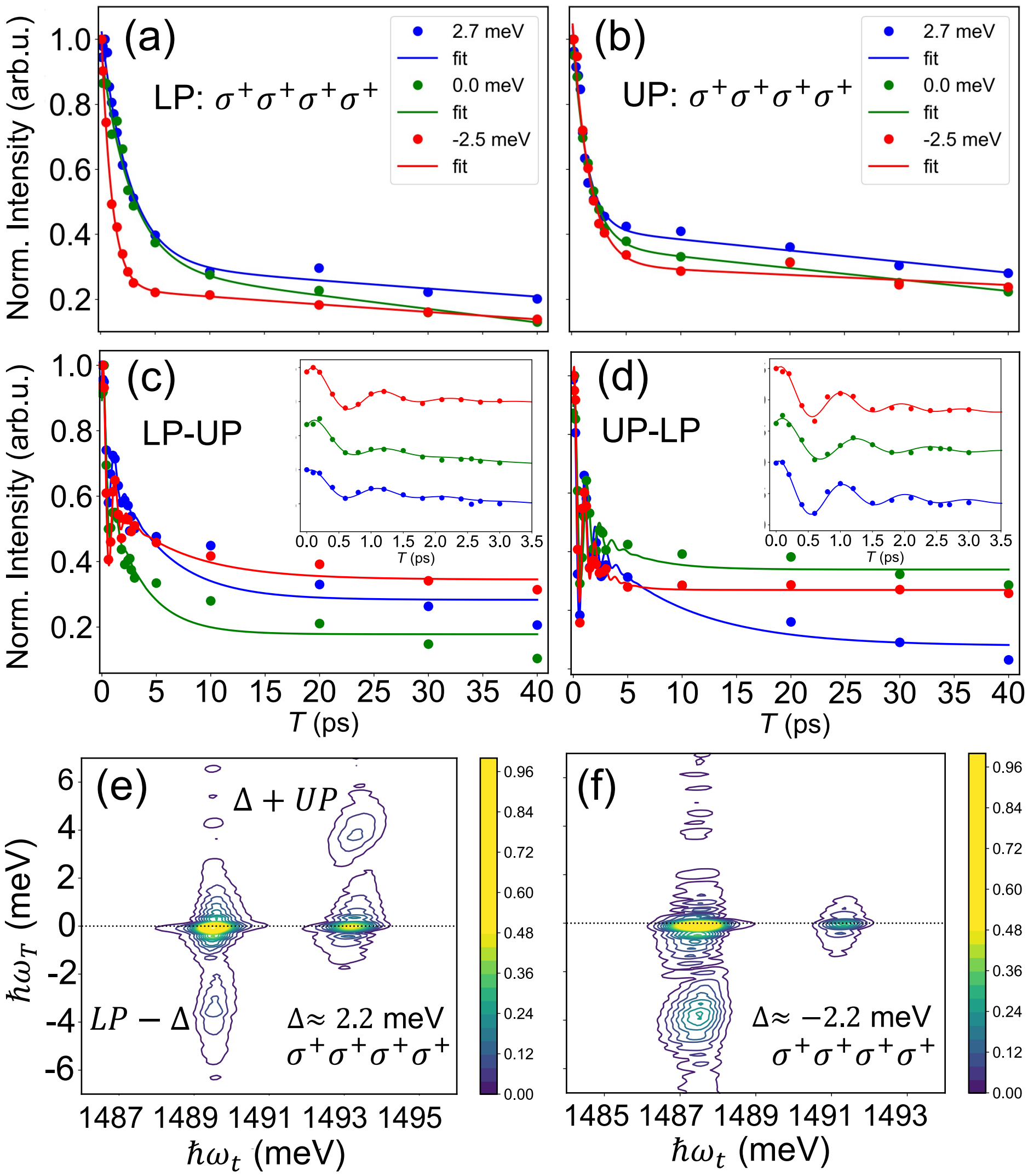

**Fig. 3**. Population time decay of the (a) diagonal lower-polariton, LP, (b) diagonal upper-polariton, UP, and off-diagonal (c) LP-UP and UP-LP features for three cavity detuning values. Short-time dependences are shown in the insets where transients are vertically spaced for clarity. Two-dimensional zero-quantum spectra, $|S_{\mathrm{I}}(\tau, \omega_T, \omega_t)|$, are shown for co-circular polarization and detuning values (e) Δ = 2.2 meV and (f) Δ = –2.2 meV.

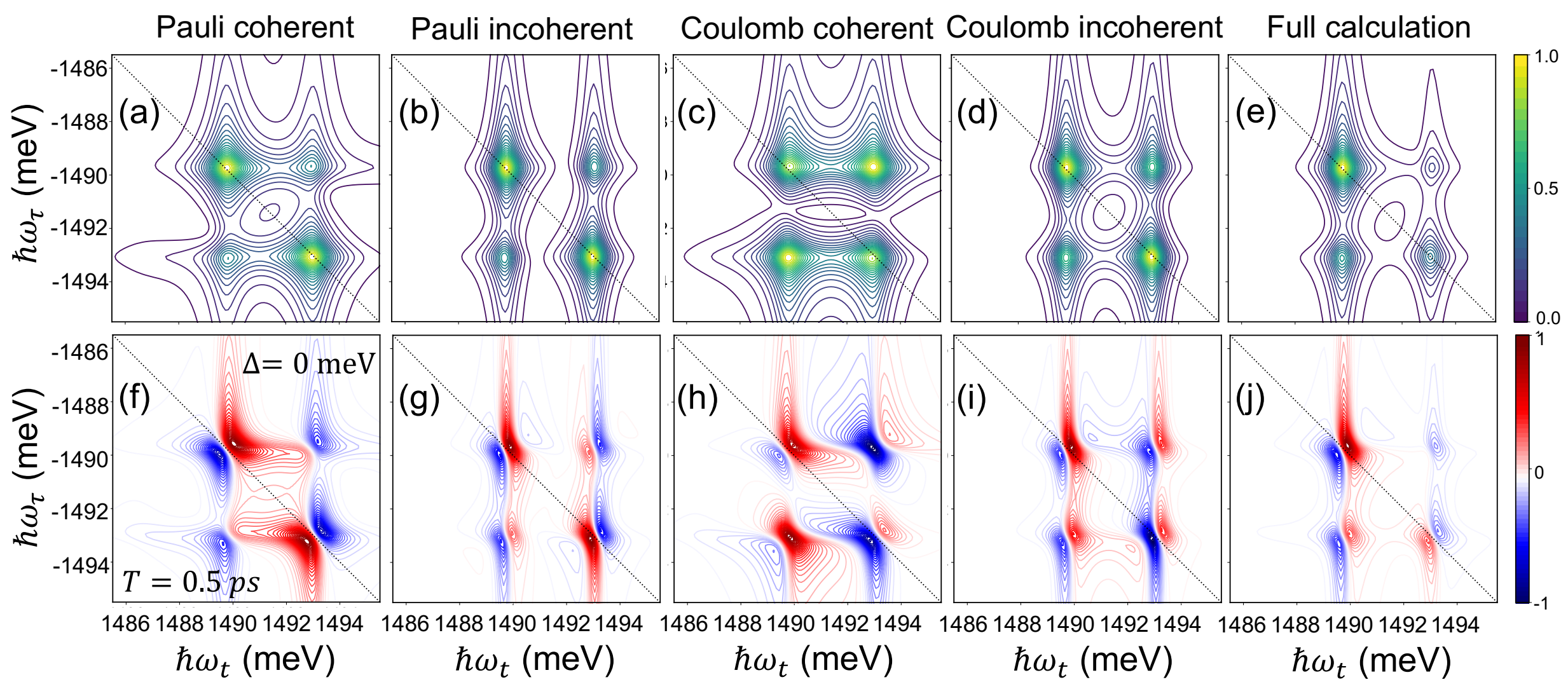

**Fig. 4**. Pauli blocking and Coulomb interaction contributions and full calculation of the simulated rephasing spectra for $T = 0.5$ ps, $\Delta = 0$ meV and co-circular polarization. (a)-(e) amplitude and (f)-(j) real parts of the spectra.

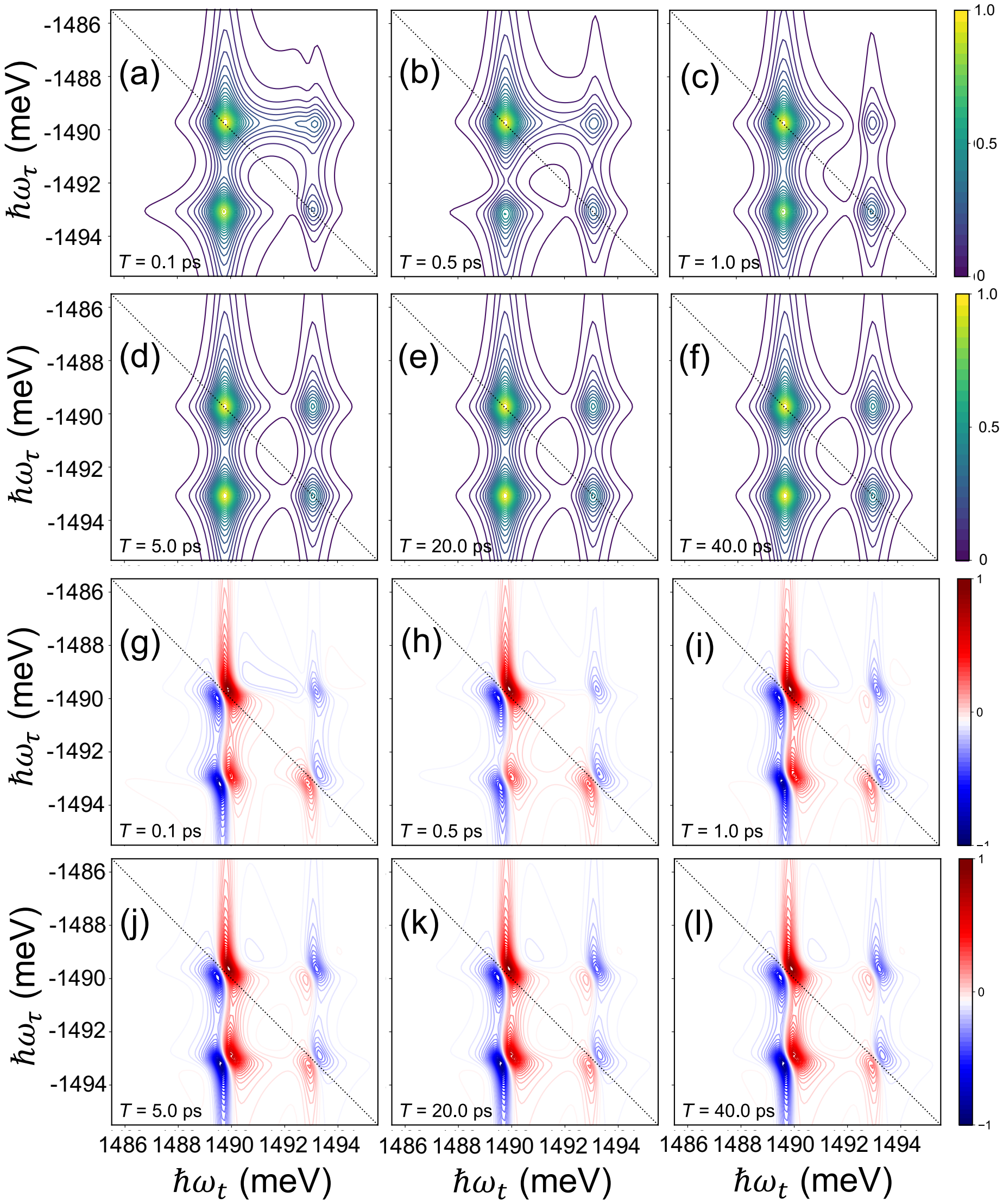

**Fig. 5**. Simulated two-dimensional rephasing spectra for the range 0.1 ps < $T$ < 40 ps for co-circular polarization and Δ = 0 meV. (a)-(f) Magnitude spectra. (g)-(l) Real part spectra.

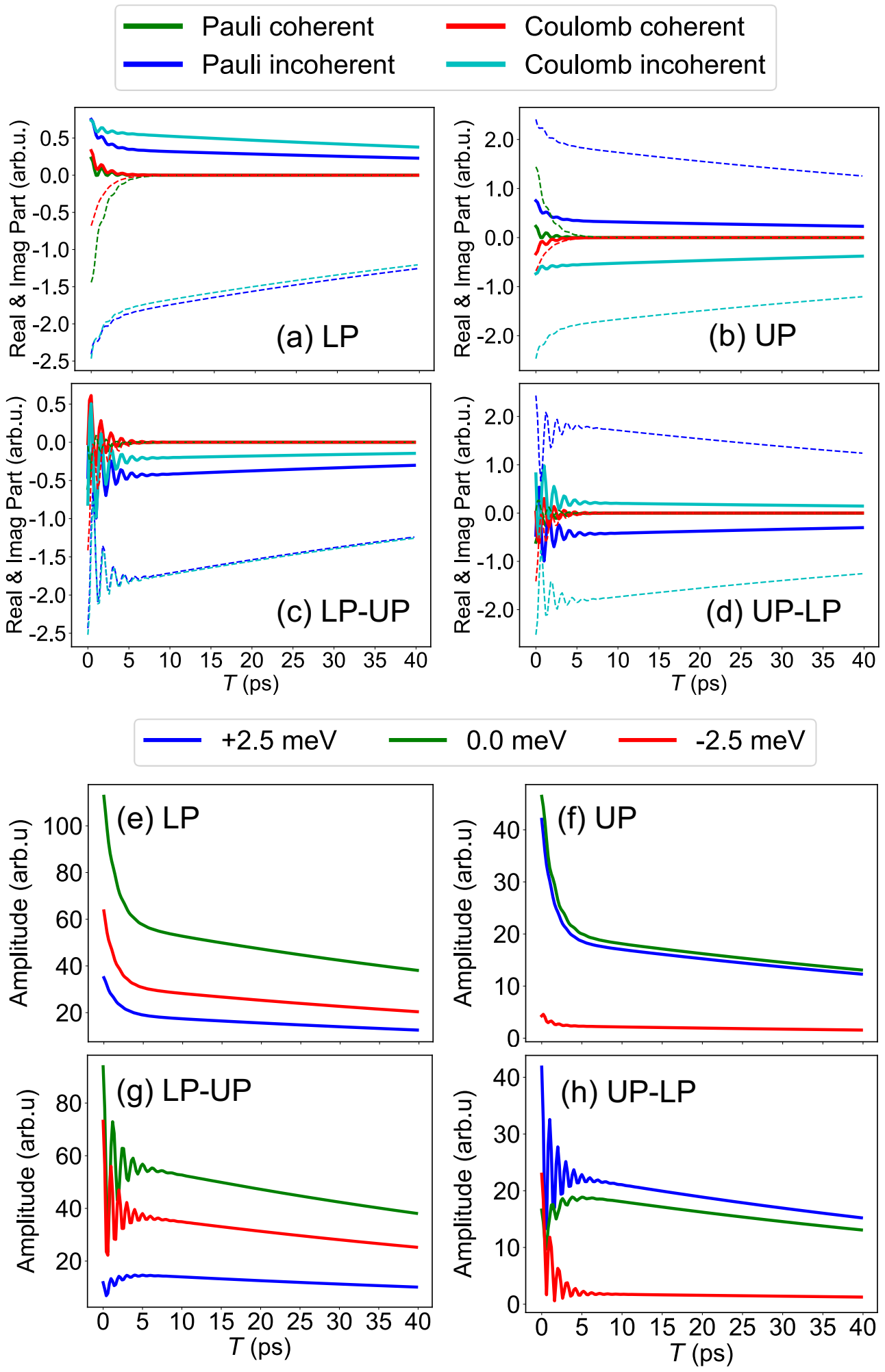

**Fig. 6**. Simulated $T$ dependences of the four main features (LP, UP, LP-UP and UP-LP) in the rephasing spectra. (a)-(d) Pauli and Coulomb contributions to the full calculation at $\Delta = 0$ meV. The real (solid lines) and imaginary (dashed lines) part of the amplitude is plotted as a function of $T$. (e)-(h) The full calculation for $\Delta = -2.2$ meV, 0 meV and 2.2 meV.

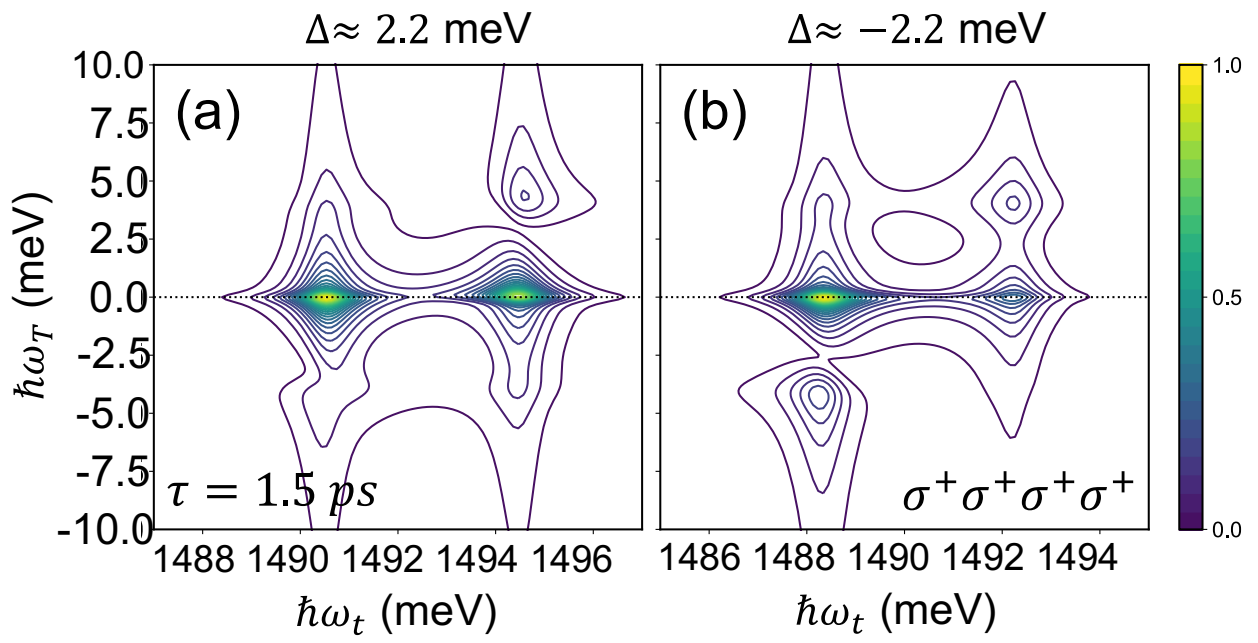

**Fig. 7**. Simulated zero-quantum spectra, $|S_{\mathrm{I}}(\tau = 1.5\ \mathrm{ps}, \omega_T, \omega_t)|$, for (a) $\Delta = 2.2$ meV and (b) $\Delta = -2.2$ meV.